\documentclass[twocolumn,showpacs,preprintnumbers,amsmath,amssymb,prl]{revtex4}

\usepackage{graphicx}
\usepackage{dcolumn}
\usepackage{bm}
\usepackage{tabularx}
\usepackage{amsmath}

\begin{document}

\title{Single-gap $s$-wave superconductivity in Cu$_x$TiSe$_2$}

\author{S. Y. Li,$^1$ Louis Taillefer,$^{1,2,*}$ G. Wu,$^3$ and X. H. Chen$^3$}

\affiliation{$^1$D{\'e}partement de physique and RQMP,
Universit{\'e} de Sherbrooke, Sherbrooke, Canada\\
$^2$Canadian Institute for Advanced Research, Toronto, Canada\\
$^3$Hefei National Laboratory for Physical Science at Microscale and
Department of Physics, University of Science and Technology of
China, Hefei, Anhui 230026, P. R. China}

\date{\today}

\begin{abstract}
The in-plane thermal conductivity $\kappa$ of the layered
superconductor Cu$_x$TiSe$_2$ was measured down to temperatures as
low as $T_c$/40, at $x$ = 0.06 near where the CDW order vanishes.
The absence of a residual linear term at $ T \rightarrow 0$ is
strong evidence for conventional $s$-wave superconductivity in this
system. This is further supported by the slow magnetic field
dependence, also consistent with a single gap, of uniform magnitude
across the Fermi surface. Comparison with the closely related
material NbSe$_2$, where the superconducting gap is three times
larger on the Nb 4$d$ band than on the Se 4$p$ band, suggests that
in Cu$_{0.06}$TiSe$_2$ the Se 4$p$ band is below the Fermi level and
the Ti 3$d$ band is alone responsible for the superconductivity.
\end{abstract}


\maketitle

Superconductivity has recently been found in the layered compound
TiSe$_2$, when its charge density wave (CDW) transition is
continuously suppressed by Cu doping \cite{Morosan1}. The phase
diagram of Cu$_x$TiSe$_2$ is illustrated in Fig. 1, where
superconductivity (SC) emerges near $x$ = 0.04 and reaches a maximum
$T_c$ of 4.15 K at $x$ = 0.08, beyond which $T_c$ decreases
\cite{Morosan1}. Such a phase diagram is reminiscent of high-$T_c$
cuprates and some heavy fermion (HF) superconductors, in which SC
appears close to where magnetic order disappears with doping or
pressure. For comparison, the phase diagram of HF superconductor
CeIn$_3$ \cite{Mathur} is reproduced in Fig. 1.

The fact that superconductivity emerges precisely at the quantum
critical point (QCP) where antiferromagnetic order vanishes in
CeIn$_3$ and CePd$_2$Si$_2$ has been viewed as a compelling argument
that this superconductivity is mediated by magnetic fluctuations
\cite{Mathur}. Similarly, valence fluctuations were suggested to
mediate the superconductivity in CeCu$_2$(Si$_{1-x}$Ge$_x$)$_2$, a
HF system with a valence transition induced by pressure \cite{Yuan}.
Theoretically, Monthoux and Lonzarich have recently shown that
density-fluctuations can mediate superconductivity and find that a
$d$-wave order parameter is favored \cite{Monthoux}. However, in
none of the QCP-related superconductivity of HF materials is the
symmetry of the order parameter known. In this context, the
occurrence of superconductivity in Cu$_x$TiSe$_2$ near its CDW QCP
is of great interest, particularly as we are able to investigate the
symmetry of the order parameter by checking if there are nodes in
the superconducting gap.

Layered dichalcogenides MX$_2$ (M is transition metal, X = S, Se, or
Te) come in two structures: 1$T$ or 2$H$. Both 1$T$ and 2$H$
structures consist of two-dimensional X-M-X layers in which the X
atom sheets exhibit a hexagonal close-packed structure and the M
atoms are in octahedral (1$T$) or trigonal prismatic (2$H$) holes
defined by the two X sheets. While CDW order has been well studied
in both 1$T$ and 2$H$ structured MX$_2$ compounds
\cite{Wilson,Pillo,Kidd,Valla,Barnett}, superconductivity was only
seen in 2$H$ structures, such as 2$H$-TaSe$_2$, 2$H$-TaS$_2$,
2$H$-NbSe$_2$, and 2$H$-NbS$_2$ \cite{Castro}. Angle-resolved
photoemission spectroscopy (ARPES) \cite{Yokoya} and thermal
conductivity \cite{Boaknin} studies have revealed that 2$H$-NbSe$_2$
($T_c$ = 7.2 K) is a multi-band $s$-wave superconductor whereby the
gap is large on one part of the Fermi surface and much smaller (3
times) on another part. By intercalating 1$T$-TiSe$_2$ with Cu,
Cu$_x$TiSe$_2$ is the first superconducting 1$T$ structured MX$_2$
compound \cite{Morosan1}. It is interesting to compare the
superconducting ground states of 2$H$-NbSe$_2$ and
1$T$-Cu$_x$TiSe$_2$.

\begin{figure}[t]
\centering \resizebox{\columnwidth }{!}{\includegraphics{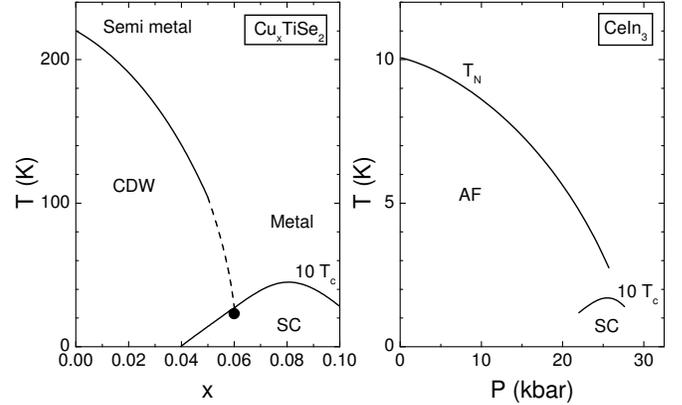}}
\caption{\label{fig1} Left: The $T-x$ electronic phase diagram of
Cu$_x$TiSe$_2$ \cite{Morosan1}. The filled circle represents the
Cu$_{0.06}$TiSe$_2$ single crystal with $T_c$ = 2.3 K for this
study. Right: Temperature-pressure phase diagram of CeIn$_3$
\cite{Mathur}.}
\end{figure}

In this Letter, we probe the superconducting ground state of
Cu$_{0.06}$TiSe$_2$ by measuring the thermal conductivity $\kappa$
of a single crystal down to 50 mK. This doping concentration is
right at the critical point where the CDW order vanishes. In zero
field, the residual linear term $\kappa_0/T$ is zero, a clear
indication that Cu$_{0.06}$TiSe$_2$ is an $s$-wave superconductor
with a gap that is finite everywhere on the Fermi surface (no
nodes). The field-dependence of $\kappa_0/T$ shows conventional
$s$-wave behavior, with no evidence for a variation of the gap
magnitude across the Fermi surface, in contrast to the case of
NbSe$_2$ where strong multi-band character is observed. The
difference is explained by examining the evolution in the band
structure of Cu$_x$TiSe$_2$ upon Cu doping.

Single crystals of Cu$_{0.06}$TiSe$_2$ were grown by the
vapor-transport technique \cite{Wu}. The copper concentration was
determined by inductively coupled plasma spectrometer (ICP) chemical
analysis, and confirmed by $c$ lattice parameter calibration
\cite{Morosan1}. The sample was cut to a rectangular shape of
dimensions 1.5 $\times$ 1.0 mm$^2$ in the plane, with 30 $\mu$m
thickness along the $c$-axis. Contacts were made directly on the
fresh sample surfaces with silver paint, which were used for both
thermal conductivity and resistivity measurements. The typical
contact resistance was 20 m$\Omega$ at low temperature. In-plane
thermal conductivity was measured in a dilution refrigerator down to
50 mK using a standard one heater-two thermometer steady-state
technique. Magnetic fields were applied along the $c$-axis and
perpendicular to the heat current.

\begin{figure}[t]
\centering \resizebox{\columnwidth }{!}{\includegraphics{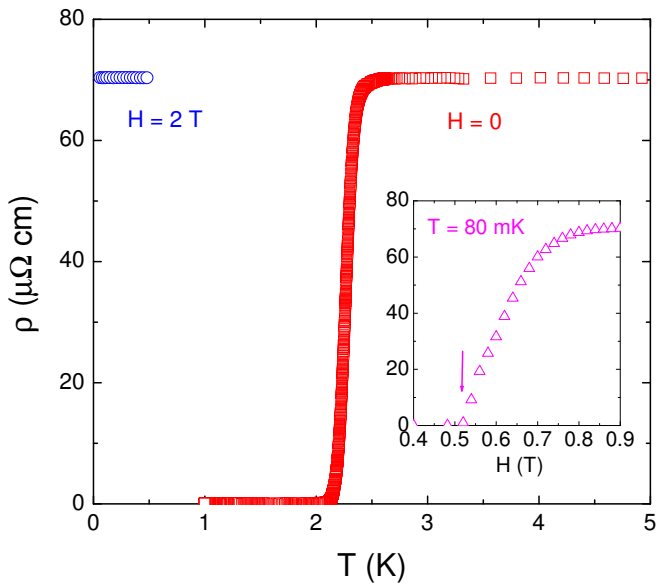}}
\caption{\label{fig2} In-plane resistivity for our
Cu$_{0.06}$TiSe$_2$ single crystal in zero field and $H$ = 2 T
applied along the $c$-axis. {\it Inset}: Field dependence of $\rho$
at $T$ = 80 mK, from which $H_{c2}$(80 mK) $\approx$ 0.52 T is
obtained (arrow).}
\end{figure}

Fig. 2 shows the in-plane resistivity of our Cu$_{0.06}$TiSe$_2$
single crystal in $H$ = 0 and 2 T. In zero field, the middle point
of the resistive transition is at $T_c$ = 2.3 K, in good agreement
with previous studies \cite{Morosan1}. The 10-90\% width of the
resistive transition is 0.15 K, indicating the high homogeneity of
our crystal. The normal-state resistivity in $H$ = 2 T is
essentially temperature independent below 0.5 K, which gives the
residual resistivity $\rho_0$ = 70.4 $\mu \Omega$ cm. This value is
comparable to that of a Cu$_{0.07}$TiSe$_2$ single crystal
\cite{Morosan2}. The inset of Fig. 2 shows the field dependence of
$\rho$ at 80 mK, from which the upper critical field $H_{c2}$(80 mK)
$\approx$ 0.52 T is obtained. The magnetoresistance is negligible
above $H$ = 1 T.

\begin{figure}[t]
\centering \resizebox{\columnwidth }{!}{\includegraphics{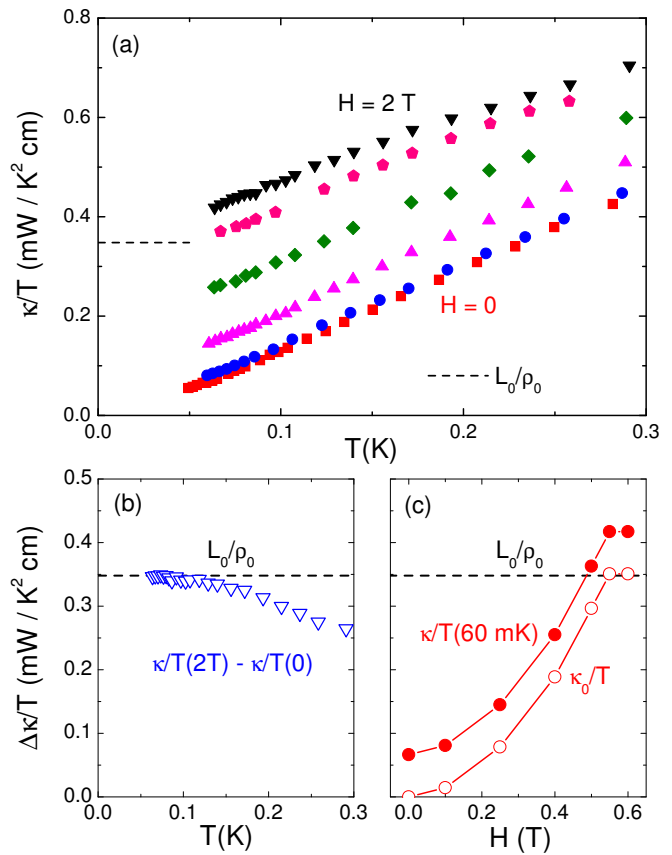}}
\caption{\label{fig3} (a) Low-temperature thermal conductivity of
Cu$_{0.06}$TiSe$_2$ in magnetic fields applied along the $c$-axis
($H$ = 0, 0.1, 0.25, 0.4, 0.5, 2 T from bottom to top). The dashed
line is the normal state Wiedemann-Franz law expectation at $T
\rightarrow 0$, namely $L_0$/$\rho_0$, with $L_0$ the Lorenz number
2.45 $\times 10^{-8}$ W $\Omega$ K$^{-2}$. (b) Difference in thermal
conductivity between $H$ = 2 T (normal state) and zero field. (c)
Field dependence of $\kappa_0 / T$. The raw data of $\kappa/T$ at 60
mK is also shown.}
\end{figure}

The in-plane thermal conductivity of Cu$_{0.06}$TiSe$_2$ is plotted
in Fig. 3a, as $\kappa / T$ vs $T$. By applying a magnetic field ($H
> H_{c1} \sim$ 5 mT \cite{Morosan2}), a roughly rigid shift
develops from the $H$ = 0 curve and eventually saturates above $H$ =
0.55 T. The measured conductivity is the sum of two contributions,
respectively from electrons and phonons, so that $\kappa = \kappa_e
+ \kappa_p$. In order to extract $\kappa_e$ at $T \rightarrow 0$ we
extrapolate $\kappa/T$ to $T = 0$, i.e. obtain the residual linear
term $\kappa_0/T$. This can be done by fitting the data to $\kappa/T
= a + bT^{\alpha-1}$, below 150 mK. In the normal state of $H$ = 2
T, this gives $\kappa_0/T$ = 0.355 $\pm$ 0.016 mW K$^{-2}$
cm$^{-1}$, with $\alpha$ = 2.24. This satisfies the Wiedemann-Franz
law, $\kappa_0/T$ = $L_0/\rho_0$, within 4\%, which validates our
method of extrapolating to $T = 0$. The same fit to the zero field
data yields a negligible residual linear term, $\kappa_0/T$ = 0.001
$\pm$ 0.004 mW K$^{-2}$ cm$^{-1}$, with $\alpha$ = 2.27.

We can obtain a value for $\kappa_0/T$ at $H = 0$ without having
recourse to any fitting or extrapolation, simply by assuming that
the Wiedemann-Franz law is perfectly obeyed in the normal state (at
2 T). Given that this law is universally obeyed in all good metals,
this is a very reasonable assumption. In Fig. 3b, the difference
$\Delta\kappa/T = \kappa/T$(2T) - $\kappa/T(0)$ is plotted. The
curve saturates below about 120 mK, to a value precisely equal to
$L_0/\rho_0$ (with $\rho_0$ measured on the same crystal with the
same contacts). Given our assumption that $\kappa_e/T$(2T) =
$L_0/\rho_0$, this implies that $\kappa_e/T(0) = 0$. The fact that
$\Delta\kappa/T$ is constant below 120 mK means that $\kappa_p$ is
the same in the superconducting state ($H$ = 0) and normal state
($H$ = 2 T). This indicates that electron-phonon scattering is weak
in this bad metal, and it only becomes significant as $T$ is
increased (accounting for the drop in $\Delta\kappa/T$ at high $T$).
By plotting $\Delta\kappa/T = \kappa/T$(2T) - $\kappa/T(H)$ and
using the same analysis, $\kappa_0/T$ in fields between 0 and 2 T is
obtained. The field dependence of $\kappa_0/T$ is plotted in Fig. 3c
(from which we see that the bulk $H_{c2}$(0) $\approx$ 0.55 T). The
raw data of $\kappa/T$ at 60 mK is also shown.

{\it Order parameter symmetry.} --- The fact that there is no
residual linear term in this layered conductor, i.e. that
$\kappa_0/T = 0$, directly implies that there are no nodes in the
gap -- the gap is non-zero everywhere on the Fermi surface. This is
strong evidence in favour of an order parameter with $s$-wave
symmetry. For unconventional superconductors with nodes in the
superconducting gap, the nodal quasiparticles will contribute a
finite $\kappa_0/T$ in zero field. For example, $\kappa_0/T$ = 1.41
mW K$^{-2}$ cm$^{-1}$ for the overdoped cuprate Tl2201, a $d$-wave
superconductor with $T_c$ = 15 K \cite{Proust}, and $\kappa_0/T$ =
17 mW K$^{-2}$ cm$^{-1}$ for the ruthenate Sr$_2$RuO$_4$, a $p$-wave
superconductor with $T_c$ = 1.5 K \cite{Suzuki}. The size of
$\kappa_0/T$ is determined by the ratio of quasiparticle velocities
parallel ($v_\Delta$) and perpendicular ($v_F$) to the Fermi surface
near the nodes \cite{Graf,Durst}. For a two dimensional $d$-wave
superconductor with a gap maximum $\Delta_0$ and a density of $n$
planes per unit cell of height $c$ one gets \cite{Durst}
\begin{equation}
\frac{\kappa_0}{T} \simeq \frac{{k_B}^2}{6}
\frac{n}{c}\kappa_F\frac{v_F}{\Delta_0}
\end{equation}
assuming $v_F \gg v_\Delta$, where $k_F$ is the Fermi wavevector.
Applying this formula to cuprate superconductors works
quantitatively very well \cite{Chiao,Hawthorn}. Using values
appropriate for Cu$_x$TiSe$_2$, namely $v_F$ = 0.4 eV\AA\
\cite{Qian}, $k_F \sim$ 0.5 \AA\ \cite{Qian}, and $\Delta_0$ = 2.14
$k_BT_c$ = 0.51 meV, Eq. 1 gives $\kappa_0/T$ = 2.4 mW K$^{-2}$
cm$^{-1}$. This estimate is several orders of magnitude larger than
the error bar on the measured $\kappa_0/T$. Therefore we can rule
out unconventional superconductivity with nodes in
Cu$_{0.06}$TiSe$_2$. Instead, we find that the superconducting state
of Cu$_x$TiSe$_2$ is characterized by a fully gapped excitation
spectrum, which most likely implies an order parameter of $s$-wave
symmetry.

\begin{figure}[t]
\centering \resizebox{\columnwidth }{!}{\includegraphics{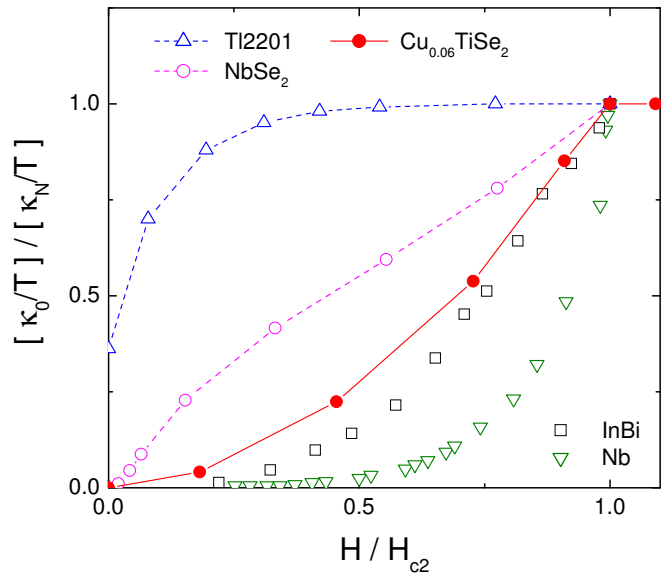}}
\caption{\label{fig4} Normalized residual linear term $\kappa_0/T$
of Cu$_{0.06}$TiSe$_2$ plotted as a function of $H/H_{c2}$. For
comparison, similar data are shown for the clean $s$-wave
superconductor Nb \cite{Lowell}, the dirty $s$-wave superconducting
alloy InBi \cite{Willis}, the multi-band $s$-wave superconductor
NbSe$_2$ \cite{Boaknin} and an overdoped sample of the $d$-wave
superconductor Tl-2201 \cite{Proust}.}
\end{figure}

{\it Single-gap superconductivity.} ---  In Fig. 4, the normalized
$\kappa_0/T$ of Cu$_{0.06}$TiSe$_2$ is plotted as a function of
$H/H_{c2}$, together with similar low-temperature data for the clean
$s$-wave superconductor Nb \cite{Lowell}, the dirty $s$-wave
superconducting alloy InBi \cite{Willis}, the multi-band $s$-wave
superconductor NbSe$_2$ \cite{Boaknin} and an overdoped sample of
the $d$-wave superconductor Tl-2201 \cite{Proust}. For a clean
type-II $s$-wave superconductor with a single gap, $\kappa$ should
grow exponentially with field (above $H_{c1}$), as is indeed
observed in Nb \cite{Lowell}. For InBi, the curve is exponential at
low $H$, crossing over to a roughly linear behaviour closer to
$H_{c2}$ as expected for $s$-wave superconductors in the dirty limit
\cite{Caroli}. A very similar behaviour was found recently in the
layered superconductor C$_6$Yb, whose $s$-wave gap was confirmed by
penetration depth measurements (see Ref. \cite{Sutherland}).
However, the thermal conductivity of multi-band $s$-wave
superconductor NbSe$_2$ \cite{Boaknin} shows a distinctly different
field dependence. In the multiband scenario, gaps of different
magnitudes are associated with different bands. Applying a field
rapidly delocalizes quasiparticle states confined within the
vortices associated with the smaller gap band, while those states
associated with the larger gap band delocalize more slowly. This
gives rise to the rapid increase in $\kappa$ at low fields
\cite{Kusonose} evident in the NbSe$_2$ data \cite{Boaknin} and in
the archetypal multi-band superconductor MgB$_2$ \cite{Sologubenko}.
Quantitatively, at $H = H_{c2}(0)/9$, $\kappa_0/T$ in NbSe$_2$ has
already risen to 1/4 of its normal-state value, while it is still
negligible in a single-gap $s$-wave superconductor. This was
explained in terms of a gap $\Delta_0(T \rightarrow 0)$ whose
magnitude on one part of the Fermi surface is 3 times smaller than
elsewhere (given that $H_{c2}(0) \propto \Delta_0^2$).

From Fig. 4, it is clear that Cu$_{0.06}$TiSe$_2$ is different from
the multi-band superconductor NbSe$_2$, and more likely a dirty
single-gap $s$-wave superconductor such as InBi. The BCS coherence
length $\xi_0 \sim$ 290 \AA\ has been estimated from the Fermi
velocity \cite{Qian}. To check if Cu$_{0.06}$TiSe2 is indeed in the
dirty limit $l < \xi_0$, we calculate the electronic mean free path
$l$ using the normal state thermal conductivity $\kappa_N$, specific
heat $c$, and Fermi velocity $v_F$. Since $\kappa_N = (1/3)cv_Fl$,
we get $l = 3(\kappa_N/T )/(\gamma v_F)$, where $\gamma = c/T$ is
the linear specific heat coefficient. With $\kappa_N/T$ = 0.355 mW
K$^{-2}$ cm$^{-1}$, $\gamma$ = 3.5 mJ mol$^{-1} $K$^{-2}$
\cite{Morosan1}, and $v_F$ = 0.4 eV\AA\ \cite{Qian}, we obtain $l
\sim 19$ \AA, one order of magnitude smaller than $\xi_0$. This
confirms Cu$_{0.06}$TiSe$_2$ to be in the dirty limit.

To explain the difference between the single-gap superconductivity
of 1$T$-Cu$_x$TiSe$_2$ and the multi-band superconductivity of
2$H$-NbSe$_2$, let us compare their electronic band structures.
High-resolution ARPES measurements on 2$H$-NbSe$_2$ \cite{Yokoya}
have shown two groups of Fermi surface (FS) sheets: a small,
holelike FS centered at the $\Gamma$ point derived from the Se 4$p$
band and larger hexagonal FS sheets around $\Gamma$(A) and K(H)
points derived from Nb 4$d$ bands. Two slightly different
superconducting gaps were found on the Nb 4$d$ FS sheets, $\Delta$ =
1.0 and 0.9 meV respectively, while no gap was detected on the Se
4$p$ FS sheet \cite{Yokoya}, at the relatively high measurement
temperature of 5.3 K = 0.74 $T_c$, in agreement with the
interpretation of the $\kappa$ data mentioned above. It is believed
that the density of states and electron-phonon coupling in the Se
4$p$ band are both smaller than in the Nb 4$d$ bands, which explains
the difference in the magnitude of the superconducting gaps on
different Fermi surfaces of 2$H$-NbSe$_2$.

Previously the Se 4$p$ band in pure 1$T$-TiSe$_2$ has been shown to
be slightly unoccupied around $\Gamma$ \cite{Pillo}, thus a small
hole pocket as in 2$H$-NbSe$_2$. The Ti 3$d$ band is only thermally
occupied at room temperature and considerably shifts towards the
occupied range upon cooling \cite{Pillo}. Very recently two groups
have reported systematic ARPES studies of 1$T$-Cu$_x$TiSe$_2$
\cite{Qian,Zhao} to explain the ``competition" between CDW and
superconductivity. Zhao {\it et al.} \cite{Zhao} found that, with Cu
doping, the electrons will fill in the Se 4$p$ hole pocket quickly
and most of them are filled in the narrow Ti 3$d$ band. Since the Se
4$p$ band is fully occupied in 1$T$-Cu$_{0.06}$TiSe$_2$, there is
only Ti 3$d$ band responsible for superconductivity. This explains
why 1$T$-Cu$_{0.06}$TiSe$_2$ is a single-gap superconductor,
different from the multi-band superconductor 2$H$-NbSe$_2$.

Based on their study, Zhao et al. \cite{Zhao} concluded that the
apparent ``competition" between CDW and superconductivity in
1$T$-Cu$_x$TiSe$_2$ is very likely a coincidence, as the doping will
increase the density of states (enhance SC) and raise the chemical
potential (suppress CDW according to the excitonic mechanism
originally proposed by Kohn \cite{Kohn}) simultaneously. And the
drop of superconductivity at high doping might be due to strong
scattering caused by the dopants. In this sense, it may not come as
a surprise that we observe conventional superconductivity in this
system. In other words, the appearence of superconductivity may have
little to do with the suppression of CDW order and associated
fluctuations.

In summary, we have used thermal conductivity to clearly demonstrate
single-gap $s$-wave superconductivity in Cu$_{0.06}$TiSe$_2$. This
rules out unconventional superconductivity with gap nodes in
Cu$_x$TiSe$_2$, despite the quantum phase transition from CDW order
to superconductivity induced by Cu doping. In contrast to the
multi-band $s$-wave superconductor NbSe$_2$, our result implies that
the Se 4$p$ band in Cu$_{0.06}$TiSe$_2$ is below the Fermi level and
only the Ti 3$d$ band is responsible for superconductivity, in
agreement with ARPES studies.

We are grateful to P. Fournier for sample characterization. This
research was supported by NSERC of Canada, a Canada Research Chair
(L.T.), and the Canadian Institute for Advanced Research. The work
in China was supported by a grant from the Natural Science
Foundation of China.\\

$^*$ E-mail: louis.taillefer@usherbrooke.ca

\end{document}